\documentclass[11pt, a4paper,sort&compress]{article} % JCAP

\usepackage{jcappub}
\usepackage{natbib}
\usepackage{cancel}
\usepackage{pifont} 
\usepackage{amsthm}         
\usepackage{amsmath} 	   
\usepackage{amssymb}       
\usepackage{amsfonts}
\usepackage{graphicx} 	    
\usepackage{verbatim}
\usepackage{color}
\usepackage{colortbl}
\usepackage{float}
\usepackage{multirow}
\usepackage{booktabs}
\usepackage{makecell}

\title{Phantom-Crossing Dark Energy and the $\Omega_m$ Tug-of-War}

\author[a]{David Shlivko}
\author[b]{and Vivian Poulin}

\affiliation[a]{Department of Physics, Princeton University, Princeton, NJ 08544, USA}
\affiliation[b]{Laboratoire univers et particules de Montpellier (LUPM),
Centre national de la recherche scientifique (CNRS) et Universit\'e de Montpellier,
Place Eug\`ene Bataillon, 34095 Montpellier C\'edex 05, France}

\emailAdd{dshlivko@princeton.edu}
\emailAdd{vivian.poulin@umontpellier.fr}

\abstract{
Recent analyses combining data from the cosmic microwave background (CMB), baryon acoustic oscillations (BAO), and Type Ia supernovae (SN) have revealed a tentative observational preference for phantom crossing in the dark energy equation of state $w$. We argue that this preference is a natural consequence of the $\Omega_m$ tensions that arise when these datasets are individually fit to $\Lambda$CDM, specifically because of the ordering $\Omega_m^\mathrm{BAO} < \Omega_m^\mathrm{CMB} < \Omega_m^\mathrm{SN}$. We show both theoretically and empirically that models with phantom crossing can shift all of these inferred $\Omega_m$ values toward mutual alignment. In contrast, quintessence theories restricted to $w \geq -1$ can alleviate the tensions with SN data but only at the cost of exacerbating the BAO-CMB discrepancy. We therefore conclude that it is the BAO and CMB measurements---not the SN data---that drive the preference for phantom crossing over quintessence in joint analyses. Moreover, we point out that SN data exhibit greater tensions with the other datasets when fit to phantom-crossing models than when fit to quintessence, causing the preference for phantom crossing to be weaker in joint CMB+BAO+SN analyses than in analyses of CMB+BAO data alone. 
}
\arxivnumber{}

\begin{document}
\notoc \maketitle
\flushbottom

\section{Introduction}
Recent measurements of the cosmic microwave background (CMB), baryon acoustic oscillations (BAO), and Type Ia supernovae (SN) have begun to hint at an observational preference for models of dynamical dark energy over a cosmological constant \cite{SN:pantheon, SN:union, popovic_dark_2025, SN:DES_collab, DESI:2024vi, desi_cosmo_2025, mustapha_throne}. In the best-fit models, the dark energy equation of state transitions from an early-time phantom regime ($w < -1$) to a late-time quintessence-like regime ($w > -1$) near redshift $z \approx 0.4$ \cite{lodha_extended_2024, wolf_scant_2024, lodha_extended_2025, gialamas_quintessence_2025, ozulker_dark_2025}. This ``phantom-crossing'' behavior is indicative of dynamics in the dark sector that extend beyond simple quintessence theories, which predict $w \geq -1$ at all times \cite{Ratra:1987rm,Peebles:1987ek,Wetterich:1987fm,frieman_pngb_1995, Coble:1996te,Turner:1997npq,Caldwell:1997ii}. Scenarios in which one observes either a real or an apparent phantom-crossing transition include theories with non-canonical kinetic terms \cite{chen_quintessential_2025, goldstein_monodromic_2025, koutroulis_uv-complete_2026}, non-minimal gravitational couplings \cite{wolf_assessing_2025, wolf_cosmological_2025, brax_weinbergs_2025, adam_comparing_2025, lopez_non-minimally_2025, nojiri_phantom_2025, nojiri_apparent_2026}, non-standard dark matter evolution \cite{yang_probing_2025, kumar_evidence_2025, giani_matter_2025, chen_evolving_2025}, and interacting dark sectors \cite{chen_quintessential_2025, andriot_phantom_2025, khoury_apparent_2025, bedroya_evolving_2025, toomey_kinetic_2025, giani_novel_2025}.

The significant theoretical implications of phantom crossing warrant a detailed examination of where exactly quintessence theories are failing to fit the data. In this work, we show that the preference for phantom crossing over quintessence is driven primarily by CMB and BAO data---partially due to better fits to the individual datasets, but also in large part due to an improvement in the consistency between them. In particular, we find that the $\Omega_m^\mathrm{BAO} < \Omega_m^\mathrm{CMB}$ discrepancy that arises when fitting these datasets to $\Lambda$CDM \cite{desi_cosmo_2025} is worsened when the data are fit to quintessence theories, whereas this tension can be alleviated in phantom-crossing models of dark energy, leading to a significantly improved $\chi^2$ in joint CMB+BAO analyses. We provide analytical arguments explaining why this behavior is expected given the observables and redshift ranges probed by these datasets.

In contrast, we find that measurements of SN distance moduli---which probe the $z \approx 0.4$ universe most precisely---\emph{weaken} the preference for phantom crossing over quintessence when included in joint analyses. Unlike the CMB-BAO tension, we show that the $\Omega_m^\mathrm{BAO} < \Omega_m^\mathrm{SN}$ discrepancy that arises within $\Lambda$CDM can be alleviated in both phantom-crossing models and quintessence scenarios, boosting the performance of both types of models over $\Lambda$CDM. However, we find that standard phantom-crossing models exhibit mild-to-moderate tensions between SN data and CMB+BAO constraints, causing their boost in performance to be somewhat diminished. These results help to explain the findings of previous studies, which have empirically shown that evidence for phantom crossing (compared to non-phantom dynamical dark energy) is stronger in joint CMB+BAO analyses than for other combinations of data \cite{sousa-neto_symbolic_2025, keeley_preference_2025, wolf_scant_2024, lodha_extended_2024, lodha_extended_2025}. 

The primary goal of this work is to clarify the distinct roles of individual datasets and of inter-dataset tensions in constraining the nature of dark energy, especially as it relates to the recent preference for phantom crossing over quintessence. Our purpose is not to provide a new model-selection analysis, which has previously been explored using both frequentist and Bayesian approaches in, e.g., \cite{SN:pantheon, SN:union, popovic_dark_2025, SN:DES_collab, DESI:2024vi, desi_cosmo_2025, mustapha_throne, lodha_extended_2025, shlivko_optimal_2025, alestas_desi_2025, cline_simple_2025, akrami_has_2025, de_souza_thawing_2025, payeur_observations_2025, park_updated_2025, keeley_could_2025, shlivko_thawing_2026}. Rather, we aim to identify the observational origin of the preference for phantom crossing independently of its extent.
In section \ref{s_models}, we define the phenomenological CPL and Pad\'e-$w$ parameterizations that will be used to model phantom crossing and quintessence theories of dark energy, and we outline the datasets and computational methods that will be used to analyze these models statistically. 
In section \ref{s_cmb}, we review the constraints on dynamical dark energy imposed by CMB data alone, including the so-called ``mirage'' of $w = -1$. In section \ref{s_bao}, we focus on BAO measurements and explain why their tension with the CMB can be relieved in phantom-crossing models of dark energy but not in quintessence theories. In section \ref{s_sn}, we discuss SN measurements, their tensions with the other datasets in $\Lambda$CDM, and the new tensions that arise in CPL models of dark energy. Finally, we conclude with a summary of our results and an outlook on the possible impact of future observational and theoretical advances in section \ref{s_conclusions}. 

\section{Models, data, and methodology}\label{s_models}
To model a broad range of dark energy dynamics capable of phantom-crossing behavior, we use the Chevallier-Polarski-Linder (CPL) parameterization \cite{chevallier_accelerating_2001,Linder:2002et,de_putter_calibrating_2008} of the equation of state, 
\begin{equation}\label{e_cpl}
w_\mathrm{CPL}(z)= w_0 + w_a z /(1+z).
\end{equation}
Meanwhile, to model quintessence theories with $w \geq -1$, we use the Pad\'e-$w$ parameterization from ref. \cite{shlivko_optimal_2025}:
\begin{equation}\label{e_pade}
	w_\text{Pad\'e}(z) = \frac{2\epsilon_0}{3 + \eta_0 (z^3+3z^2+3z)} - 1.
\end{equation} 
For positive $\epsilon_0$ and $\eta_0$, this parameterization provides a highly precise model of ``thawing'' quintessence dynamics \cite{caldwell_limits_2005, linder_paths_2006, cahn_field_2008}, in which the dark energy equation of state is restricted to $w > -1$ and decreases asymptotically toward $w = -1$ as the redshift $z$ increases. This behavior is representative of a broad and well-motivated class of quintessence theories, and it is aligned with the observational indications of an equation of state that increases over time. The parameters in this model represent the present-day equation of state, 
\begin{equation}
	\epsilon_0 = \frac{3}{2}(1+w_\text{Pad\'e}(0)),
\end{equation}
and its rate of change, 
\begin{equation}
	\eta_0 = -\frac{d\ln(1+w_\text{pad\'e})}{d\ln(1+z)}|_{z=0}.
\end{equation}

We perform statistical comparisons between $\Lambda$CDM and the CPL and Pad\'e-$w$ models of dark energy using likelihoods from the following datasets: 

\begin{enumerate}
	\item \textbf{CMB.} 
	We use the \texttt{P-ACT} combination of power spectra from the Planck satellite and ACT, which includes low-$\ell$ TT data from the Planck PR3 likelihood \cite{planck_2020_like}, low-$\ell$ EE data from the Sroll2 likelihood \cite{pagano_reionization_2020}, partial high-$\ell$ data from Planck PR3 restricted to $\ell < 1000$ in TT and $\ell < 600$ in TE/EE, and the ACT DR6 dataset \cite{naess_atacama_2025, louis_atacama_2025}.
	We also include a combination of CMB lensing data from ACT \cite{ACT_lensing1, ACT_lensing2, ACT_lensing3} and Planck's PR4 (NPIPE) maps \cite{carron_Planck_lensing}, choosing the \texttt{actplanck\_baseline} likelihood variant.
	\item \textbf{BAO.} We use the DESI DR2 observations of galaxies, quasars, and the Lyman-alpha forest \cite{desi_forest_2025, desi_cosmo_2025, andrade_validation_2025, casas_validation_2025, brodzeller_construction_2025}, measuring the comoving angular diameter distance $D_M(z)$, the Hubble distance $D_H(z)$, or the angle-averaged quantity $D_V(z) \equiv (zD_M(z)^2D_H(z))^{1/3}$ in seven redshift bins. The individual DESI observations span $0.1 \leq z \leq 4.2$, while the binned effective redshifts range from $z = 0.295$ to $z = 2.330$. 
	\item \textbf{SN.} We use three different supernova datasets in this work. The PantheonPlus compilation \cite{SN:pantheon} contains 1701 measurements of 1550  Type Ia supernovae from redshifts $0.001 < z < 2.26$. (Note that the 111 measurements at $z \leq 0.01$ are filtered out in statistical analyses.) The Union3 compilation \cite{SN:union} groups 2087 Type Ia supernova measurements from redshifts $0.01 < z < 2.26$ into twenty-two effective redshift bins ranging from $z = 0.05$ to $z = 2.26$. The DES-Dovekie compilation \cite{popovic_dark_2025} comprising 1623 high-redshift ($0.1 < z < 1.15$) and 197 low-redshift ($0.025 < z < 0.1$) Type Ia supernovae. We respectively denote these three supernova datasets by SN$_\mathrm{P}$, SN$_\mathrm{U}$, and SN$_\mathrm{D}$. 
\end{enumerate}

Our analyses are performed within \texttt{cobaya} \cite{cobaya1, cobaya2}, making use of the \texttt{CAMB} Boltzmann solver \cite{camb1, camb2}. Within \texttt{CAMB}, we use the parameterized post-Friedmann approach \cite{fang_ppf} to compute dark energy perturbations, and we modify the 2016 \texttt{HMcode} algorithm \cite{mead_hmcode_2015, mead_accurate_2016} for computing non-linear matter power spectra to use the (tabulated) Pad\'e-$w$ parameterization input into \texttt{CAMB} instead of the default CPL parameterization.
For accurate estimates of best-fit chi-squared values, we run six parallel iterations of the \texttt{iminuit} minimizer \cite{iminuit}. Posterior densities are computed using the Metropolis-Hastings MCMC sampler with dragging \cite{metropolis1, metropolis2, metropolis_drag} and plotted with \texttt{GetDist} \cite{getdist}. The convergence of MCMC chains is determined using the Gelman-Rubin statistic \cite{gelman_inference_1992}, for which we require $R - 1 < 0.01$. Estimates of tensions between datasets are derived from these chains using non-Gaussian kernel density estimation methods included in the \texttt{Tensiometer} package \cite{tensiometer}.

\begin{table}[t]
\centering
	\begin{tabular}{|l|l|l|}
		\hline
		Parameter & Prior (incl. CMB) & Prior (no CMB)\\
		\hline
		$\ln(10^{10}A_s)$ & $\mathcal{U}[2.9, 3.2]$ & $\mathcal{U}[1.61, 3.91]$\\
		$n_s$ & $\mathcal{U}[0.9, 1.05]$& $\mathcal{U}[0.8, 1.2]$\\
		$\tau$ & $\mathcal{U}[0.02, 0.1]$&$\mathcal{U}[0.02, 0.1]$\\
		$100\theta_\text{MC}$ & $\mathcal{U}[1.03, 1.05]$&-\\
		$\omega_b$ & $\mathcal{U}[0.02, 0.025]$&derived\\
		$\omega_c$ & $\mathcal{U}[0.1, 0.14]$&derived\\
		$H_0$ (km/s/Mpc) & derived & $\mathcal{U}[20, 100]$ \\
		$\Omega_b$ & derived & $\mathcal{U}[0.03, 0.07]$ \\
		$\Omega_m$ & derived & $\mathcal{U}[0.1, 0.9]$ \\
		$w_0$ & $\mathcal{U}[-3, 1]$ & $\mathcal{U}[-3, 1]$ \\
		$w_a$ & $\mathcal{U}[-10, 2]$ &$\mathcal{U}[-10, 2]$ \\
		$\sqrt{\epsilon_0}$ & $\mathcal{U}[0, \sqrt{30}]$ &$\mathcal{U}[0, \sqrt{30}]$ \\
		$\eta_0$ & $\mathcal{I}[0, 1000]$ & $\mathcal{I}[0, 1000]$\\
		\hline
	\end{tabular}
	\caption{Parameters used in our statistical analyses and their priors. Different parameters and priors are used depending on whether CMB likelihoods are included in the analysis. Uniform distributions are denoted by $\mathcal{U}[a, b]$. $A_s$ is the amplitude of the primordial power spectrum, $n_s$ is the spectral index, $\tau$ is the optical depth, and $\theta_\text{MC}$ represents the angular size of the sound horizon at recombination. The matter fraction $\Omega_m$ includes contributions from both baryons (b) and cold dark matter (c), whose physical densities are defined as $\omega_i \equiv \Omega_i h^2$, with $h \equiv H_0 / (100\text{ km/s/Mpc})$.  %The variables $A_\text{ACT}$ and $P_\text{ACT}$ correspond to ACT's dipole and polarization calibration parameters. 
	Finally, the dark energy equation of state is defined by either the CPL parameters $\{w_0, w_a\}$ or the Pad\'e-$w$ parameters $\{\epsilon_0, \eta_0\}$ per eqs. (\ref{e_cpl}-\ref{e_pade}). The prior on $\eta_0$ (denoted $\mathcal{I}$) is taken to be the informed prior derived in ref. \cite{shlivko_thawing_2026}.}\label{t_priors}
\end{table}

Throughout this work, we assume a spatially flat universe, a single massive neutrino with $m_\nu = 0.06$ eV, and $N_\text{eff} = 3.044$ total neutrino species. The priors for variable parameters in our analyses are shown in table \ref{t_priors}; these differ depending on whether CMB likelihoods are included in a given analysis. For the Pad\'e-$w$ parameters $\epsilon_0$ and $\eta_0$, we use the informed priors from ref. \cite{shlivko_thawing_2026}; this choice allows for a theoretically grounded visualization of posterior densities in the Pad\'e-$w$ parameter space, and it does not affect the key results of this work that draw on comparisons of best-fit $\chi^2$ values.

\section{The role of CMB data}\label{s_cmb}

Modern measurements of the cosmic microwave background are independently consistent with $\Lambda$CDM \cite{planck_2020_like, louis_atacama_2025}. 
Additionally, the power spectrum of anisotropies in the CMB provides a tight constraint on the present-day matter density $\rho_{m,0}$ independent of the nature of dark energy \cite{act_extended_2025}. Within $\Lambda$CDM, this  allows the precisely measured angular acoustic scale, interpreted as an inferred distance to the last scattering surface at $z^* \approx 1090$,
%distance to the last scattering surface (corresponding to redshift $z^* \approx 1090$),
\begin{equation}\label{e_lss}
	D_\mathrm{LSS} = \int_0^{z^*}\frac{dz}{H(z)} \propto  \int_0^{z^*}\frac{dz}{\sqrt{\rho_{m,0}(1+z)^3 + \rho_{DE,0}}},
\end{equation}
 to provide a similarly tight constraint \cite{louis_atacama_2025} on the matter fraction 
 \begin{equation}
 	\Omega_m^\mathrm{CMB} = \frac{\rho_{m,0}}{\rho_{m,0}+\rho_{DE,0}} = 0.312 \pm 0.007 \quad \text{[$\Lambda$CDM: ACT+Planck]}.
 \end{equation}
 (Note that we neglect the contribution of radiation to the total energy density when integrating over matter- and dark-energy-dominated epochs of the universe.)
 
In models of dark energy with an evolving equation of state $w(z)$, the distance to last scattering can be written in the more general form
  \begin{equation}\label{e_lss2}
	D_\mathrm{LSS} \propto \int_0^{z^*}\frac{dz}{\sqrt{\rho_{m,0}(1+z)^3 + \rho_{DE,0}\ e^{3\int_0^z \frac{1+w(z')}{1+z'}dz'}}}.
\end{equation}
Hypothetically, if one were to impose the same value of $\Omega_m \approx 0.31$ in these dynamical models, then maintaining an observationally compatible prediction for $D_\mathrm{LSS}$ would require $w(z)$ to ``average out,'' loosely speaking, to its $\Lambda$CDM limit of $w = -1$. For the family of CPL models in particular, this requires a crossing of the phantom divide at $w = -1$ near redshifts $z \approx 0.4$ probed by late-time measurements, a phenomenon known as the mirage of $w = -1$ \cite{linder_mirage_2007}. 
 
\begin{figure*}[t]
	\includegraphics[width=0.99\textwidth]{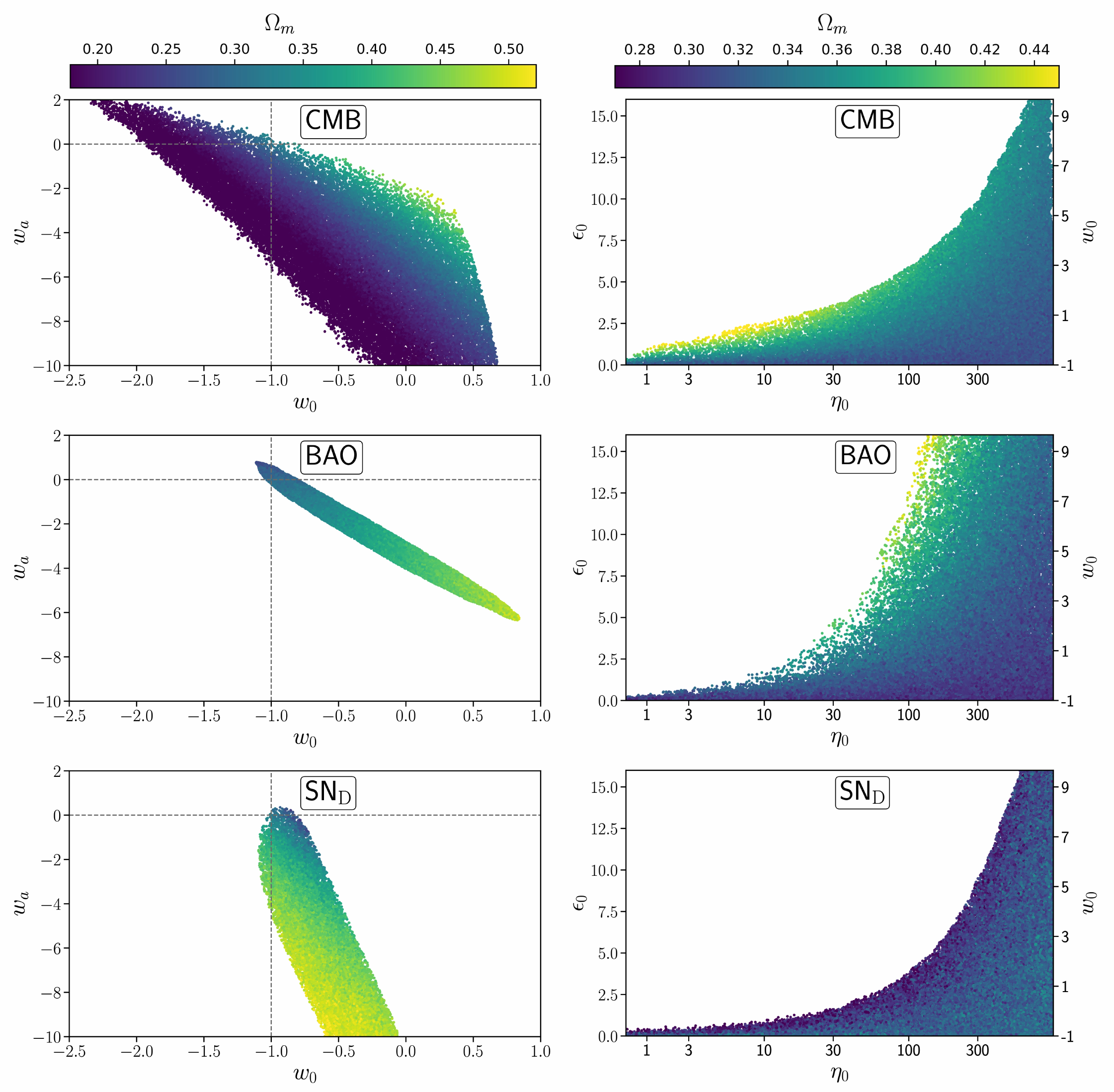}
	\caption{\label{f_colored}Observational constraints on the CPL (left) and Pad\'e-$w$ (right) parameter spaces, colored in accordance with observationally compatible values of $\Omega_m$ (colorbars are shared within columns). The top row shows constraints from CMB data, the middle row shows constraints from BAO data, and the bottom row shows constraints from SN data (using the DES-Dovekie supernova sample SN$_\mathrm{D}$). The individual plotted points represent MCMC samples with at least 5\% of the maximum 2D posterior density (which generalizes the $2\sigma$ region of a Gaussian distribution). Dashed lines are drawn to cross at the $\Lambda$CDM limit of the CPL parameter space; the $\Lambda$CDM limit corresponds to $\epsilon_0 = 0$ for Pad\'e-$w$.}
\end{figure*}

 In practice, however, dynamical dark energy theories can allow for significant deviations from $\Omega_m \approx 0.31$. In the more general case where $\Delta\Omega_m \equiv \Omega_m - \Omega_m^{\Lambda\mathrm{CDM}} \neq 0$, the requirement to cross the phantom divide at $w=-1$ is replaced by the crossing of a new ``pivot value'' $w_p$ at a similar redshift of $z \approx 0.4$. For small $\Delta \Omega_m \lesssim 0.1$, linearizing the expression for $D_\mathrm{LSS}$ around $w_p = -1$ and $\Omega_m \approx 0.3$ gives the approximation \cite{huterer_probing_2001, frieman_probing_2003, linder_mirage_2007}
 \begin{equation}
 	w_p \approx -1 + 3.6\Delta\Omega_m.
 \end{equation}
This means that CMB constraints on CPL parameters are expected to shift toward higher values of $w_0$ and/or $w_a$ when $\Omega_m \gtrsim 0.31$ and vice versa. This also means that quintessence theories satisfying $w(z) > -1$ at all times can still be compatible with CMB data, as long as $\Omega_m \gtrsim 0.31$. These expectations are consistent with the constraints from full CMB likelihoods shown in the top row of figure \ref{f_colored}.

In addition to the distance to last scattering, CMB data are sensitive to the dynamics of dark energy through the reconstructed lensing power spectrum, through the effects of gravitational lensing on two-point anisotropy spectra at high $\ell$, and in principle (though much more weakly) through the late-time integrated Sachs-Wolfe effect at low $\ell$. These effects can help to discriminate between scenarios in which $w$ and $\Omega_m$ are both higher or both lower. %, since these predict different histories of structure formation.  
Current CMB data give a slight edge to phantom and phantom-crossing models over $\Lambda$CDM, but only with a marginal $\Delta\chi^2 = -2.7$. Since quintessence theories lie on the opposite side of the $\Lambda$CDM boundary, they generically provide worse fits to CMB data than $\Lambda$CDM ($\Delta\chi^2 \geq 0$). Nevertheless, many Pad\'e-$w$ models (namely, those plotted in the top-right panel of figure \ref{f_colored}) still fall well within observational uncertainties. 

\section{The role of BAO data}\label{s_bao}

Much like the CMB data, BAO measurements from DESI DR2 are independently compatible with $\Lambda$CDM. The fit is slightly better for both Pad\'e-$w$ ($\Delta\chi^2 \approx -2.3$) and CPL models ($\Delta \chi^2 \approx -4.7$), but these improvements are still below $2\sigma$ in significance. 
Within $\Lambda$CDM, however, the best-fit value of $\Omega_m^\mathrm{BAO} = 0.297 \pm 0.009$ is noticeably lower than that inferred from the CMB \cite{desi_cosmo_2025}. 
This discrepancy amounts to a $1.3\sigma$ tension (or $1.7\sigma$ if one considers the joint $\Omega_m$-$H_0r_d$ parameter space) that amplifies the preference for CPL models (which can relieve this tension) in joint CMB+BAO analyses.

\begin{figure}[t]
\begin{center}
	\includegraphics[width=0.85\textwidth]{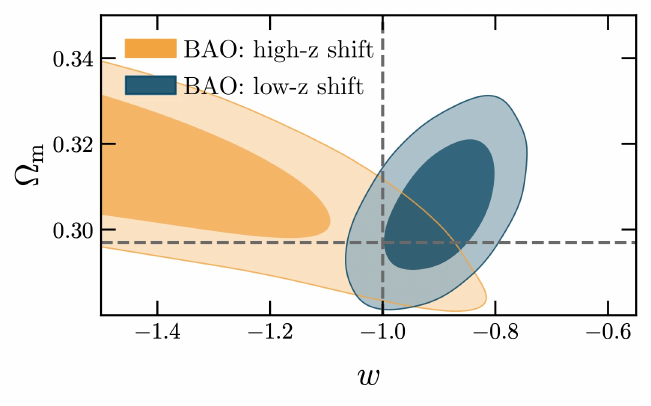}
	\caption{MCMC constraints from DESI DR2 BAO on two toy models, in which $w(z)$ is a piecewise-constant function that is allowed to deviate from $w = -1$ in either the ``low-$z$'' regime ($z < 0.4$) or the ``high-$z$'' regime ($z > 0.4$). Dashed lines indicate the $\Lambda$CDM limit with DESI's best-fit  $\Omega_m^\mathrm{BAO} = 0.297$. The inferred value of $\Omega_m$ increases as $w$ increases in the low-$z$ regime, but the reverse is true as $w$ increases in the high-$z$ regime.}\label{f_toy_mcmc}
	\end{center}\vspace{-0.4em}
\end{figure} 

Unlike phantom-crossing models, thawing quintessence models cannot relieve the CMB-BAO tension. At best, they leave the tension unchanged, and in many cases, they exacerbate it. As a result, in joint CMB+BAO analyses, the Pad\'e-$w$ models that would have improved the fit to BAO data alone become disfavored, leading to a trivial net improvement $\Delta\chi^2_\mathrm{CMB+BAO} = -0.2$ over $\Lambda$CDM. Meanwhile, CPL models can simultaneously improve the fit to CMB data, improve the fit to BAO data, and improve the concordance between them, yielding a net $\Delta\chi^2_\mathrm{CMB+BAO} = -7.6$ exceeding the sum of improvements in fit to each individual dataset. This is a relatively robust $2.3\sigma$ result.

We will now briefly explain \emph{why} the CMB-BAO tension can be alleviated in phantom-crossing models of dark energy but not in simpler theories of quintessence. 
To begin, notice that the correlation between $w(z)$ and $\Omega_m$ is qualitatively different in BAO inference (middle row of figure \ref{f_colored}) than in CMB inference (top row). This is because, unlike $D_\mathrm{LSS}$, BAO observables respond differently to changes in $w$ depending on whether those changes occur at low redshifts ($z \lesssim 0.4$) or at higher redshifts ($z \gtrsim 0.4$). We illustrate these correlations using piecewise-constant toy models in figure \ref{f_toy_mcmc}, and we show how they can be understood analytically in Appendix \ref{a_obs}.

\begin{figure}[t!]
\begin{center}
	\includegraphics[width=0.9\textwidth]{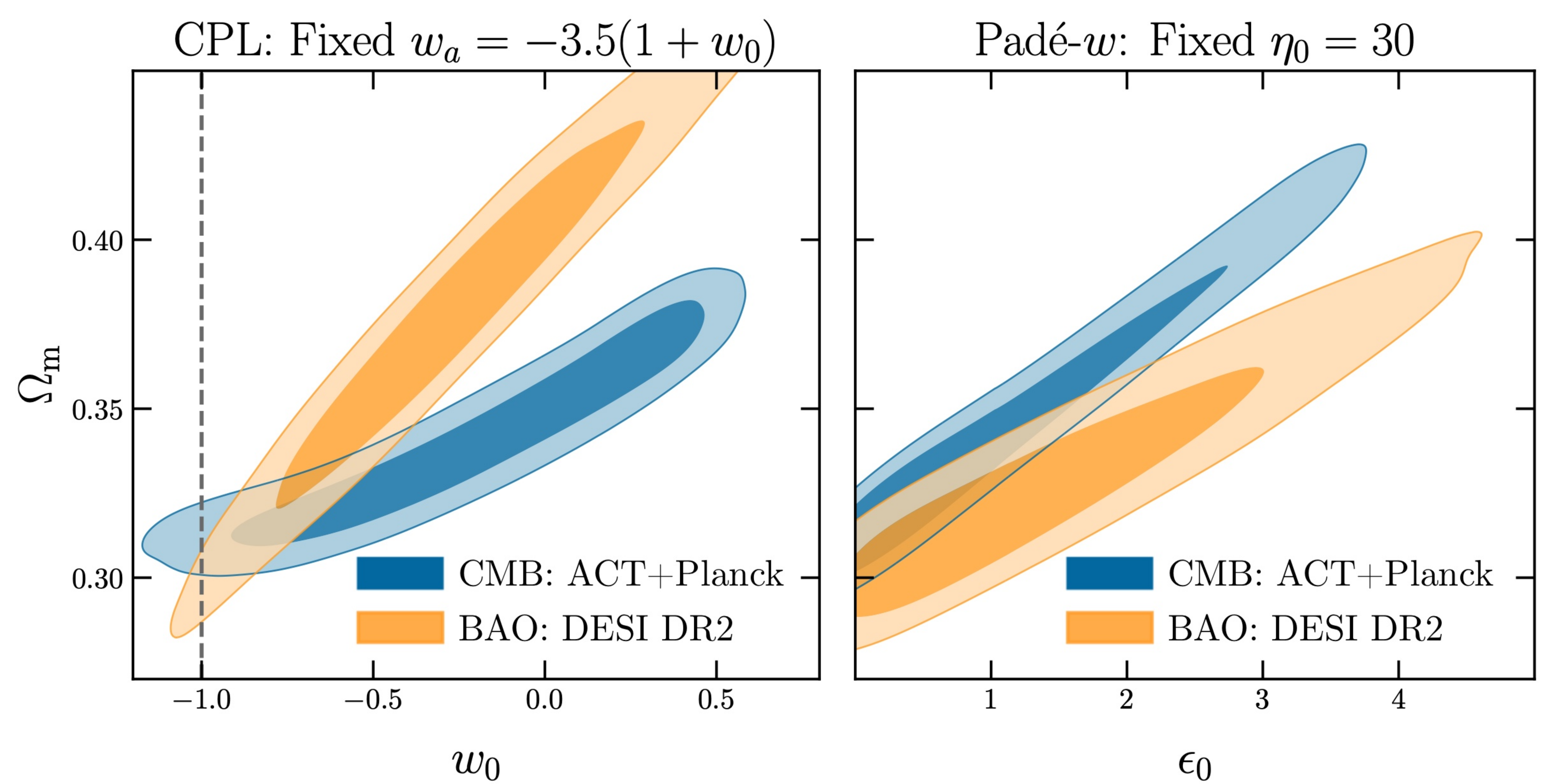}
	\caption{CMB- and BAO-based correlations between $\Omega_m$ and either the CPL parameter $w_0$ (left panel) or the Pad\'e-$w$ parameter $\epsilon_0$ (right panel) controlling deviations from $\Lambda$CDM. For visual clarity, the parameters $w_a$ and $\eta_0$ have not been varied independently (see text for further explanation). The $\Lambda$CDM limit in the left panel ($w_0=-1$) is marked with a dashed line; in the right panel, this limit corresponds to $\epsilon_0 = 0$. Notice that in the (phantom-crossing) CPL fits, deviations from $\Lambda$CDM can alleviate the CMB-BAO tension in $\Omega_m$, whereas in the (thawing-quintessence) Pad\'e-$w$ fits, deviations from $\Lambda$CDM only exacerbate it.
		}\label{f_bao_cmb}
	\end{center}
\end{figure}

In phantom-crossing models, these opposite correlations lead to a compounded shift in $\Omega_m$ relative to $\Lambda$CDM. In particular, models with $w(z \lesssim 0.4) > -1$ and $w(z \gtrsim 0.4) < -1$ can significantly increase the BAO-inferred value of $\Omega_m$. Meanwhile, $\Omega_m^\mathrm{CMB}$ reacts only to the average value of $w$ across all redshifts and can therefore remain relatively close to its $\Lambda$CDM value. 
Visually, one can follow the path of increasing $\Omega_m$ in the middle-left panel of figure \ref{f_colored} while remaining closer to an iso-$\Omega_m$ contour in the top-left panel. This dynamic is further illustrated in the left panel of figure \ref{f_bao_cmb}, where we fix $w_a \equiv -3.5(1+w_0)$ to roughly lie along the BAO constraint; we can then clearly see how the increase of $\Omega_m$ with $w_0$ is more gradual for CMB inference than for BAO inference. This allows the original tension $\Omega_m^\mathrm{BAO} < \Omega_m^\mathrm{CMB}$ within $\Lambda$CDM to be alleviated where the two contours cross.

In thawing quintessence theories, on the other hand,  $\Omega_m^\mathrm{BAO}$ increases more slowly than $\Omega_m^\mathrm{CMB}$ as we deviate from $\Lambda$CDM. This is because when $w(z)$ shifts higher at \emph{all} $z$, the inferred $\Omega_m^\mathrm{BAO}$ is pulled upward by the shift at $z \lesssim 0.4$ but simultaneously pulled downward by the shift at $z \gtrsim 0.4$, whereas $\Omega_m^\mathrm{CMB}$ is pulled strictly upward by the increase in $w$. We illustrate this result in the right panel of figure \ref{f_bao_cmb} for a family of Pad\'e-$w$ models with fixed $\eta_0 = 30$, noting that this behavior is mirrored across the full range of possible $\eta_0$ (though it would be obscured if we were to marginalize over $\eta_0$). In contrast to CPL models, the CMB- and BAO-inferred values of $\Omega_m$ now \emph{diverge} with increasing $\epsilon_0$, making the CMB-BAO tension \emph{worse}, not better, than in $\Lambda$CDM. 
An observational preference for phantom crossing over quintessence is therefore to be expected in joint CMB+BAO analyses when independent fits to $\Lambda$CDM yield $\Omega_m^\mathrm{BAO} < \Omega_m^\mathrm{CMB}$.

It is worth reiterating that the panels in figure \ref{f_bao_cmb} are meant to be interpreted through the lens of deviations from $\Lambda$CDM: Pad\'e-$w$ models exacerbate the CMB-BAO tension, while phantom-crossing CPL models alleviate it. One should \emph{not} visually compare the two models against each other across panels, because the posterior shading does not share the same scale. For example, the same $\Lambda$CDM fit appears worse in the left panel (when compared to alternative CPL fits) than in the right panel (when compared to alternative Pad\'e-$w$ fits).

\section{The role of SN data}\label{s_sn}

\begin{table*}[t!]
\centering\footnotesize
\renewcommand{\arraystretch}{1.15}
\begin{tabular*}{\textwidth}{@{\extracolsep{\fill}}lcc}
\hline
\textbf{Analysis / Model}
&\textbf{Dataset / Improvement in fit ($\chi^2_\mathrm{model} - \chi^2_\mathrm{\Lambda CDM}$)} \\
\hline
\end{tabular*}
\begin{tabular*}{\textwidth}{@{\extracolsep{\fill}}lccccc}
\textbf{Individual}
& CMB
& BAO
& $\mathrm{SN_P}$
& $\mathrm{SN_U}$
& $\mathrm{SN_D}$\\
Pad\'e-$w$
& 0.0
& -2.3
& -1.0
& -2.8
&  -3.2 \\
CPL
& -2.7
& -4.7
& -0.5
& -3.4
&  -6.1 \\
\hline
\end{tabular*}
\begin{tabular*}{\textwidth}{@{\extracolsep{\fill}}lcccccc}
\textbf{Joint (i)}
& $\mathrm{CMB+SN_P}$
& $\mathrm{CMB+SN_U}$
& $\mathrm{CMB+SN_D}$
& $\mathrm{BAO+SN_P}$
& $\mathrm{BAO+SN_U}$
& $\mathrm{BAO+SN_D}$\\
Pad\'e-$w$
& -2.0
& -5.2
& -3.8
& -4.3
&  -9.4
& -7.7\\
CPL
& -2.2
& -6.8
& -4.7
& -4.9
&  -10.1
& -7.2\\
\hline
\end{tabular*}
\begin{tabular*}{\textwidth}{@{\extracolsep{\fill}}lcccc}
\textbf{Joint (ii)}
& $\mathrm{CMB+BAO}$
& $\mathrm{CMB+BAO+SN_P}$
& $\mathrm{CMB+BAO+SN_U}$
& $\mathrm{CMB+BAO+SN_D}$\\
Pad\'e-$w$
& -0.2
& -4.1
& -7.4
& -7.1\\
CPL
& -7.6
& -7.8
& -14.0
& -10.4\\
\hline
\end{tabular*}
\caption{\label{t_chi2}
Improvements in $\chi^2$ for best-fit Pad\'e-$w$ models (representing thawing quintessence) and CPL models (exhibiting phantom-crossing behavior) relative to $\Lambda$CDM, shown for individual datasets and joint analyses. Negative values indicate better fits than $\Lambda$CDM. The first set of joint analyses includes CMB+SN and BAO+SN combinations, while the second set of joint analyses includes CMB+BAO data with and without supernovae. The outperformance of CPL over Pad\'e-$w$ is most apparent in this second set of joint analyses, where the CPL models alleviate the CMB-BAO tension in $\Omega_m$.} 
\end{table*}

On their own, current SN data are generally compatible with $\Lambda$CDM. Fits to the PantheonPlus and Union3 datasets are only mildly improved by thawing quintessence and phantom-crossing CPL models, and these improvements are comparable to each other (see table \ref{t_chi2}). Improvements when fitting the DES-Dovekie dataset are slightly greater and give an edge to CPL ($\Delta\chi^2 = -6.1$) compared to Pad\'e-$w$ ($\Delta\chi^2 = -3.2$), but the statistical significance of either of these improvements over $\Lambda$CDM is still no greater than $2\sigma$.

\begin{figure*}[t!]
	\includegraphics[width=\textwidth]{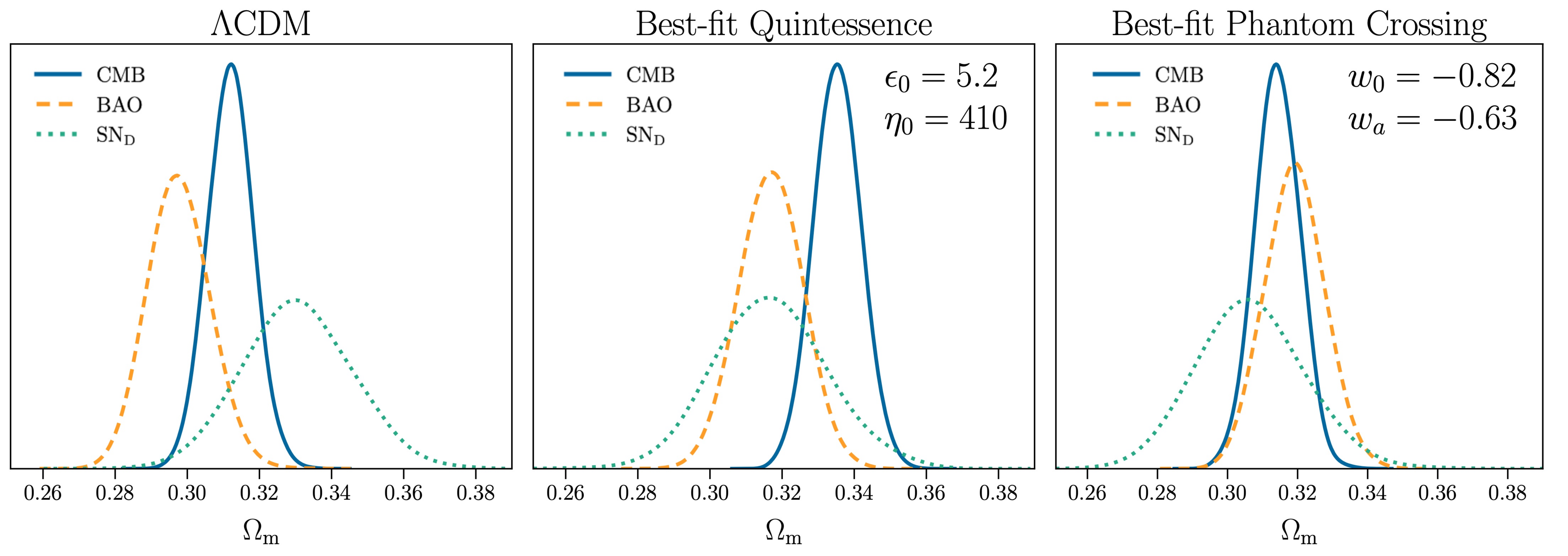}
	\caption{Marginalized posterior distributions of $\Omega_m$ in $\Lambda$CDM (left panel), in the best-fit Pad\'e-$w$ model to CMB+BAO+SN data (middle panel), and in the best-fit CPL model to CMB+BAO+SN data (right panel). For the purposes of illustration, we only show results using the DES-Dovekie supernova sample (SN$_\mathrm{D}$). Notice that the $\Lambda$CDM tension between BAO and SN data is resolved in both types of dynamical dark energy models, but the BAO-CMB tension cannot be alleviated in thawing quintessence (Pad\'e-$w$) models, since $\Omega_m^\mathrm{CMB}$ increases at least as much as $\Omega_m^\mathrm{BAO}$ (as discussed in section \ref{s_bao}). Only the phantom-crossing CPL models can bring all three inferences of $\Omega_m$  into agreement.}\label{f_omegas}
\end{figure*}

Within $\Lambda$CDM, the SN datasets prefer relatively high values of $\Omega_m$ \cite{desi_cosmo_2025}, creating tensions with both CMB data ($1.1$-$1.6\sigma$) and BAO data ($1.7$-$2.1\sigma$). 
% DES-Dovekie analysis infers a relatively high $\Omega_m^\mathrm{SN} = 0.330 \pm 0.015$, which is within $1.1\sigma$ of $\Omega_m^\mathrm{CMB}$ but exhibits a more severe $1.9\sigma$ tension with $\Omega_m^\mathrm{BAO}$ \cite{popovic_dark_2025, louis_atacama_2025, desi_cosmo_2025}. 
 Fortunately, as we derive in Appendix \ref{a_obs}, SN observables generate a consistent negative correlation between the dark energy equation of state at low redshifts and the inferred value of $\Omega_m^\mathrm{SN}$. This correlation, visualized in the bottom row of figure \ref{f_colored}, runs opposite to the positive CMB- and BAO-based correlations that we had seen previously. As a result, both quintessence and phantom-crossing models can lower $\Omega_m^\mathrm{SN}$ while simultaneously raising $\Omega_m^\mathrm{BAO}$ and $\Omega_m^\mathrm{CMB}$ relative to $\Lambda$CDM, relieving the BAO-SN and/or CMB-SN tensions, as illustrated in figure \ref{f_omegas} using the DES-Dovekie supernova sample. Note that purely phantom dark energy theories (or inverted phantom-crossing models with $w < -1$ at late times) would have the opposite effect on each $\Omega_m$ and exacerbate these tensions instead.
 
\begin{figure}[h]
	\includegraphics[width=\textwidth]{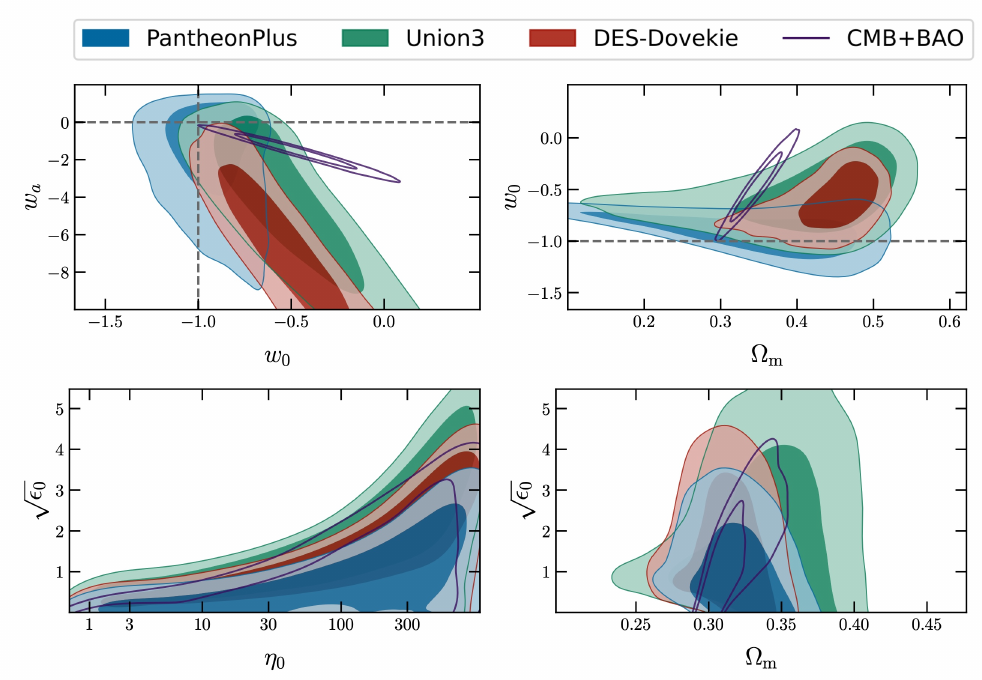}
	\caption{MCMC results showing joint 68\% and 95\% credible regions for the matter fraction $\Omega_m$ and either the CPL parameters $\{w_0, w_a\}$ (top row) or the Pad\'e-$w$ parameters $\{\epsilon_0, \eta_0\}$ (bottom row). Dashed lines are drawn to indicate the values of $w_0$ and $w_a$ in the $\Lambda$CDM limit. Supernova-based constraints are presented as shaded contours and contrasted against the (non-shaded) CMB+BAO constraints. Notice that in CPL models, SN datasets (especially DES-Dovekie) exhibit greater tensions with the CMB+BAO constraints than in the Pad\'e-$w$ parameter space.}\label{f_omw0wa_CPL}
\end{figure}

 Because quintessence and phantom-crossing models are equally capable of relieving the SN tensions, joint CMB+SN and BAO+SN analyses exhibit comparable fits when using the Pad\'e-$w$ and CPL parameterizations, as seen in the first set of joint fits in table \ref{t_chi2}. Even the slight preference for CPL that we had originally seen from the DES-Dovekie dataset is attenuated in these joint analyses, as tensions in the CPL parameter space (visualized in the top row of figure \ref{f_omw0wa_CPL}) weaken the overall goodness of fit for the optimal phantom-crossing models. Quantitatively, the tension in CPL models between DES-Dovekie and CMB+BAO data is $2.0\sigma$, and even the milder tensions when swapping in PantheonPlus supernovae ($1.2\sigma$) and Union3 supernovae ($0.8\sigma$) penalize phantom-crossing models relative to Pad\'e-$w$ fits, for which these tensions all fall below the level of $0.2\sigma$ (see the bottom row of figure \ref{f_omw0wa_CPL}).
 
 In full CMB+BAO+SN analyses, there is a residual preference for phantom-crossing CPL models over quintessence due to the improvement in CMB+BAO fit discussed in section \ref{s_bao}. As can be seen from the second set of joint fits in table \ref{t_chi2}, however, the gap in $\chi^2$ between CPL and Pad\'e-$w$ is reduced when SN data are included in the analyses. Note that this gap in $\chi^2$ is an illustrative but not formal measure of the preference for CPL over Pad\'e-$w$. Because these models are not nested, quantifying the preference for phantom crossing over quintessence is not as straightforward in a frequentist setting as quantifying their respective improvements over $\Lambda$CDM. Nevertheless, the quoted $\chi^2$ values are sufficient for the purpose of identifying the source, rather than the extent, of the preference for phantom crossing.
 
\section{Conclusions and outlook}\label{s_conclusions}

In this work, we have shown that the recent observational preference for phantom-crossing dynamics in the dark energy equation of state over a purely quintessence-like evolution with $w(z) > -1$ is driven primarily by the combination of CMB and BAO data. In part, this is because CMB and BAO measurements are individually better fit by phantom-crossing CPL models than quintessence. A significant role is also played, however, by the tension between these datasets when they are fit to $\Lambda$CDM, where one infers $\Omega_m^\mathrm{BAO} < \Omega_m^\mathrm{CMB}$. As discussed in section \ref{s_bao} and visualized in figure \ref{f_omegas}, this tension can be relieved in phantom-crossing models but not in quintessence theories, causing the former to achieve significantly better fits in joint CMB+BAO analyses. 

The current tension between CMB and BAO data in $\Lambda$CDM persists across a variety of CMB datasets \cite{desi_act_2025, lynch_whats_2025}. Additionally, this tension persists when allowing the curvature $\Omega_k$ to be nonzero\footnote{In principle, a positive $\Omega_k$ can resolve the CMB-BAO tension without invoking dynamical dark energy \cite{chen_its_2025}. However, because current CMB posteriors prefer $\Omega_k < 0$ \cite{louis_atacama_2025, desi_cosmo_2025, chen_its_2025}, the preference for CPL models over a cosmological constant (or over quintessence) is only marginally diminished in models with nonzero $\Omega_k$.} or allowing the sum of neutrino masses to be greater than the experimental lower bound of $0.06$ eV \cite{esteban_fate_2020, de_salas_2020_2021}, because in each of these cases, $\Omega_m^\mathrm{CMB}$ is driven higher within $\Lambda$CDM (see table V and figure 16 of ref. \cite{desi_cosmo_2025}). As a result, the outperformance of phantom-crossing models over quintessence is a relatively robust outcome of CMB+BAO analyses using current observational data.

In contrast, including supernova data in joint analyses serves only to diminish the overall preference for phantom-crossing dark energy models over quintessence, though the preference for each of these over $\Lambda$CDM is amplified. This is because the tensions that appear in $\Lambda$CDM between $\Omega_m^\mathrm{SN}$ and the values inferred from CMB or BAO measurements can be resolved with or without phantom crossing, as discussed in section \ref{s_sn} and visualized in figure \ref{f_omegas}. The CPL fits are somewhat degraded, however, by mild-to-moderate tensions in the ($w_0$, $w_a$, $\Omega_m$) parameter space between SN data and CMB+BAO constraints, as shown in the top row of figure \ref{f_omw0wa_CPL}. Depending on the supernova compilation, these tensions range from $0.8\sigma$ to $2.0\sigma$, compared to the $<0.2\sigma$ tensions in the Pad\'e-$w$ ($\epsilon_0$, $\eta_0$, $\Omega_m$) parameter space shown in the bottom row. When supernovae are included in joint analyses, therefore, the gap in $\chi^2$ between CPL and Pad\'e-$w$ is reduced compared to CMB+BAO analyses alone (as shown in table \ref{t_chi2}). When allowing for nonzero curvature, this gap can be reduced even further \cite{akrami_has_2025, dinda_physical_2025}. 

As future data releases offer improved statistical precision and greater control over systematic uncertainties, some or all of the inter-dataset tensions discussed in this work may evolve over time. If the CMB-BAO tension is tempered or inverted, then phantom-crossing models will lose a significant source of their advantage over quintessence. If this tension is resolved through an increase in $\Omega_m^\mathrm{BAO}$ rather than a decrease in $\Omega_m^\mathrm{CMB}$ (the latter of which could result from, e.g., a systematic correction to the measurement of $\tau$ \cite{sailer_dispuable_2026, jhaveri_turning_2025, weiner_high-redshift_2026}), then a simultaneous concordance between CMB, BAO, and SN data may even be possible within $\Lambda$CDM. On the other hand, if the CMB-BAO tension persists and future SN measurements shift to prefer CPL fits that are more closely aligned with CMB+BAO constraints, then the overall preference for phantom crossing stands to be strengthened accordingly.

It is also possible that future data releases improve precision without systematically shifting any constraints, making each of the tensions identified in this work even greater in significance. In this case, the simultaneous reconciliation of CMB, BAO, and SN data may require new theoretical advances. This could take the form of a more elaborate phantom-crossing model that relieves the CMB-BAO tension without creating new tensions with SN data. We note, however, that if discrepancies between transverse BAO measurements of $D_A(z) = (1+z)^{-1}D_M(z)$ and SN measurements of $D_L(z)=(1+z)D_M(z)$ build up at overlapping redshifts, reconciling these measurements will not be possible by simply modifying $w(z)$. In the absence of errors in the data, such a scenario could indicate a violation of the distance duality relation connecting $D_A$ and $D_L$ (see, e.g., refs. \cite{Teixeira:2025czm, afroz_hint_2025, das_cosmology-independent_2026} for recent phenomenological analyses of such violations). 

An alternative resolution to the three-way tension in $\Omega_m$ could involve modifying our theoretical description of the universe at higher redshifts \cite{weiner_high-redshift_2026}. Investigations of modified-recombination scenarios \cite{baryakhtar_cosmology_2024, smith_dynamical_2025, mirpoorian_is_2025, smith_serendipitous_2025, poulin_implications_2025} and modified pre-recombination expansion histories \cite{wang_dark_2024, chaussidon_early_2025, poulin_implications_2025, wang_dark_2025, reeves_multiprobe_2025, khalife_spt-3g_2025, garny_dark_2025} have found that the CMB-BAO tension can be at least partially alleviated through these early-time dynamics, reducing the statistical preference for phantom crossing at late times \cite{adi_lowering_2025, gonzalez-fuentes_exploring_2026}. Any residual tensions with SN data could then potentially be alleviated with a simple quintessence field. Note that these modified early-time dynamics simultaneously increase the CMB-based inference of $H_0$, diminishing the tension with measurements of the local distance ladder \cite{riess_comprehensive_2022, collaboration_local_2025, freedman_status_2025}. Because quintessence theories would otherwise exacerbate the Hubble tension by predicting lower values of $H_0$ than either $\Lambda$CDM or phantom-crossing models \cite{banerjee_hubble_2021, shlivko_thawing_2026}, they are doubly benefitted by these early-time modifications. 

In future work, assuming the observational preference for phantom-crossing dark energy persists, it will become increasingly important to measure the statistical significance of this preference. Given the non-nested relationship between CPL and precise models of thawing quintessence such as Pad\'e-$w$, a quantitative frequentist comparison may benefit from using alternative phantom-crossing parameterizations that directly generalize the dynamics of quintessence. A more theoretically grounded comparison could also be performed using Bayesian statistics once the range of microphysics of realistic phantom-crossing theories is better understood and informed priors on phantom-crossing dynamics become possible to develop. Such comparisons will further benefit from combining observations of the background evolution of the universe with measurements of clustering and growth of structure, especially in cases where the phantom crossing arises from microphysics involving dark matter (see, e.g., \cite{yang_probing_2025, kumar_evidence_2025, giani_matter_2025, chen_evolving_2025, chen_quintessential_2025, andriot_phantom_2025, khoury_apparent_2025, bedroya_evolving_2025, toomey_kinetic_2025, giani_novel_2025}).

 \vspace{0.1in}
\noindent
\textbf{Acknowledgments.} \\
%\begin{center}
%\textbf{ACKNOWLEDGMENTS}
%\end{center}
We would like to thank Alek Bedroya, Sufia Birmingham, Pedro Ferreira, Josh Frieman, Mustapha Ishak, Josh Kable, Nick Kokron, Nick Patino, Paul Steinhardt, and William Wolf for helpful conversations. We are also grateful to the organizers of the COSMO-25 conference, where collaboration on this work began. DS is supported in part by the DOE grant number DEFG02-91ER40671 and by the Simons Foundation grant number 654561. The simulations performed in this work utilized computational resources managed and supported by Princeton Research Computing, a consortium of groups including the Princeton Institute for Computational Science and Engineering (PICSciE) and the Office of Information Technology's High Performance Computing Center and Visualization Laboratory at Princeton University. VP acknowledges the European Union’s Horizon Europe research and innovation programme under the Marie Sk\l odowska-Curie Staff Exchange grant agreement No 101086085 -- ASYMMETRY.
This work received funding support from the European Research Council (ERC) under the European Union’s HORIZON-ERC-2022 (grant agreement no. 101076865).

\appendix
\section{Correlations between the dark energy equation of state and inferred values of $\Omega_m$}\label{a_obs}

Predicting the correlation between the dark energy equation of state $w(z)$ and the inferred value of $\Omega_m$ is straightforward in CMB analyses, where the data precisely constrain the distance to last scattering,
  \begin{equation}
	D_\mathrm{LSS} \propto \int_0^{z^*}\frac{dz}{\sqrt{\rho_{m,0}(1+z)^3 + \rho_{DE,0}\ e^{3\int_0^z \frac{1+w(z')}{1+z'}dz'}}},
\end{equation}
and provide an independent measure of $\rho_{m,0}$. The effect of increasing $\Omega_m$ is therefore to directly decrease $\rho_{DE,0}$ and therefore increase $D_\text{LSS}$.  On the other hand, increasing $w(z)$ at any redshift $z$ will necessarily decrease $D_\text{LSS}$. As a result, these two quantities must be positively correlated if $D_\text{LSS}$ is to remain fixed at its measured value.

The correlation between $w(z)$ and $\Omega_m$ in BAO-based inference is much more subtle. Recall that uncalibrated BAO measurements provide information about comoving angular-diameter distances $D_M(z)$ and Hubble distances $D_H(z) =H^{-1}(z)$ up to an overall multiplicative factor (representing the inverse sound horizon $r_d^{-1}$ at decoupling) \cite{desi_cosmo_2025}. Because the sound horizon itself is not directly constrained, we can model the BAO observables as describing the \emph{shapes} of $\ln D_M(z)$ and $\ln D_H(z)$ up to an overall additive constant, or equivalently, as describing the derivatives of those quantities,

\begin{equation}
\begin{aligned}
	\mathcal{O}_D(z) &\equiv \left(\frac{d\ln D_M(z)}{dz}\right)^{-1} \\ &\propto  \sqrt{\Omega_m(1+z)^3 + \Omega_\mathrm{DE}\,e^{3\int_0^z \frac{1+w(z')}{1+z'}dz'}}\int_0^{z}\frac{dz'}{\sqrt{\Omega_m(1+z')^3 + \Omega_\mathrm{DE}\,e^{3\int_0^{z'} \frac{1+w(z'')}{1+z''}dz''}}}
\end{aligned}
\end{equation}
and
\begin{equation}
	\mathcal{O}_H(z) \equiv \frac{d\ln D_H(z)}{dz} \propto \frac{\Omega_m(1+z)^2 + \Omega_\mathrm{DE} \frac{1+w(z)}{1+z}\,e^{3\int_0^{z} \frac{1+w(z')}{1+z'}dz'}}{\Omega_m(1+z)^3 + \Omega_\mathrm{DE}\,e^{3\int_0^z \frac{1+w(z')}{1+z'}dz'}}.
\end{equation}
(Note that in computing $\mathcal{O}_D(z)$, we have dropped additive constants dependent only on $z$.) Unlike the case with $D_\text{LSS}$, it is not so clear whether these observables will increase or decrease in response to shifts in $w(z)$ and $\Omega_m$. In fact, the answer depends intricately on which observable is measured, the redshift at which the observable is measured, and the redshift at which $w$ is shifted. The resulting correlation between $w(z)$ and $\Omega_m$ will therefore similarly depend on these three inputs. 

\begin{figure}[t!]
\begin{center}
	\includegraphics[width=.9\textwidth]{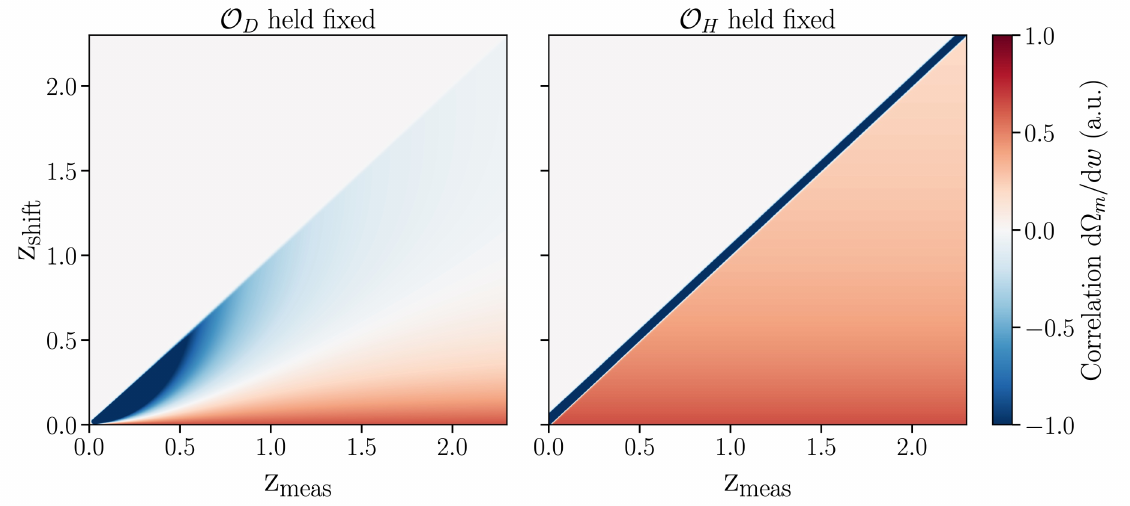}
	\caption{Correlations between $\Omega_m$ and the dark energy equation of state $w$ induced by measurements of $\mathcal{O}_D$ (uncalibrated integrated-distance observables) and $\mathcal{O}_H$ (uncalibrated Hubble-distance observables). Positive correlations are shaded red and negative correlations are shaded blue; their magnitudes are given in arbitrary units. Correlations are shown as a function of the redshift $z_\mathrm{meas}$ at which measurements are taken and as a function of the redshift $z_\mathrm{shift}$ at which $w$ is shifted away from $-1$. In the left panel with $\mathcal{O}_D$ held fixed, these shifts are modeled as delta functions. In the right panel with $\mathcal{O}_H$ held fixed, these shifts are modeled as short but finite periods of $w \neq -1$ near $z_\mathrm{shift}$ in order to illustrate the strip of strong negative correlations where $z_\mathrm{shift} = z_\mathrm{meas}$. 
	}\label{f_app_heatmap}
	\end{center}
\end{figure}

In figure \ref{f_app_heatmap}, we present two heatmaps illustrating the correlation 
\begin{equation}
	\left. \frac{d\Omega_m}{dw} \equiv -\frac{d\mathcal{O}_i(z_\mathrm{meas})/dw(z_\mathrm{shift})}{d\mathcal{O}_i(z_\mathrm{meas})/d\Omega_m}\right|_{\Omega_m = 0.3,\, w(z) = -1}
\end{equation}
assuming that the observables $\mathcal{O}_i \in \{\mathcal{O}_D, \mathcal{O}_H\}$ are fixed by measurements at a given redshift $z_\mathrm{meas}$ and that $w(z)$ shifts away from $-1$ at an independently specified redshift $z = z_\mathrm{shift}$. Notice that shifts in $w$ only affect observations (and hence the inferred value of $\Omega_m$) if $z_\mathrm{shift} \leq z_\mathrm{meas}$. We see that measurements of $\mathcal{O}_D$ can lead to a positive correlation between $w$ and $\Omega_m$ if $z_\mathrm{shift} \ll z_\mathrm{meas}$ or a negative correlation if they are similar in magnitude. Meanwhile, measurements of $\mathcal{O}_H$ lead to a positive correlation between $w$ and $\Omega_m$ for all $z_\mathrm{shift} < z_\mathrm{meas}$ but produce a strong negative correlation at $z_\mathrm{shift} = z_\mathrm{meas}$. We can also see that even high-redshift measurements of $\mathcal{O}_D$ and $\mathcal{O}_H$ remain sensitive probes of the dark energy equation of state---despite the low $\Omega_\mathrm{DE}$  at those redshifts---due to the integrated impact of $w(z)$.

\begin{figure}[t!]
\begin{center}
	\includegraphics[width=0.7\textwidth]{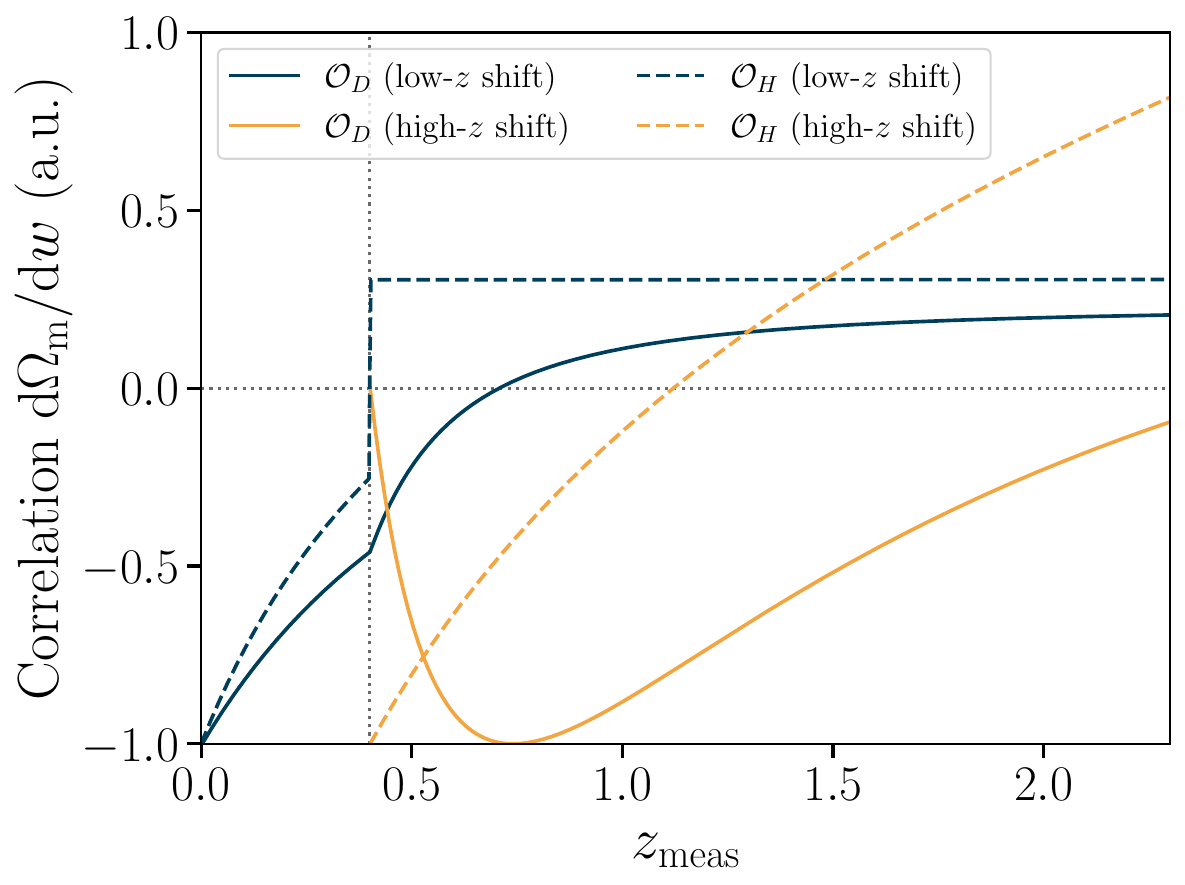}
	\caption{	Correlations between $\Omega_m$ and the dark energy equation of state $w$ induced by measurements of $\mathcal{O}_D$ (solid curves) and $\mathcal{O}_H$ (dashed curves) for two piecewise-constant models of $w(z)$. In the first model, $w$ is shifted uniformly away from $-1$ at low redshifts $z \leq 0.4$ (dark blue). In the second model, $w$ is shifted uniformly away from $-1$ at high redshifts $z \geq 0.4$ (orange). Notice that	measurements of both observables at $z_\mathrm{meas} < 0.4$ lead to negative correlations between $\Omega_m$ and $w(z < 0.4)$, whereas the sign of correlations induced by measurements at higher redshifts can vary. 
	}\label{f_toy}
	\end{center}
\end{figure}

To better understand how $\Omega_m^\mathrm{BAO}$ will shift in quintessence theories and in phantom-crossing models of dark energy relative to $\Lambda$CDM, we will now consider sustained shifts in $w(z)$ that are piecewise-constant rather than short-lived, delta-function-like deviations from $w(z) = -1$. We will analyze two scenarios: In the low-$z$ scenario, $w(z)$ is shifted uniformly away from $-1$ only in the redshift range $z \leq 0.4$; in the high-$z$ scenario, $w(z)$ is shifted uniformly away from $-1$ in the complementary redshift range $z \geq 0.4$. The resulting correlations d$\Omega_m/$d$w$ are plotted in figure \ref{f_toy} as a function of $z_\mathrm{meas}$. Solid curves correspond to fixed measurements of $\mathcal{O}_D$, while dashed curves correspond to fixed measurements of $\mathcal{O}_H$. The color of the curve distinguishes between the low-$z$ scenario (dark blue) and the high-$z$ scenario (orange), the latter of which only has an effect on the inferred $\Omega_m$ for measurements at $z_\mathrm{meas} \geq 0.4$.

Notice that in the range of redshifts probed by BAO ($0.3 \lesssim z_\text{eff} \lesssim 2.3$), the correlations induced between $\Omega_m$ and low-$z$ shifts in $w$ (blue curves) are predominantly positive. In contrast, the correlations induced between $\Omega_m$ and high-$z$ shifts in $w$ (orange curves) are predominantly negative. These analytic predictions are in line with the correlations observed empirically in the MCMC results from figure \ref{f_toy_mcmc}. 

%\newpage 
Conveniently, uncalibrated measurements of Type Ia supernovae convey information about the shape of the luminosity distance curve,
\begin{equation}
	D_L(\Omega_m, w(z) \mid z_\mathrm{meas}) \propto D_M(\Omega_m, w(z) \mid z_\mathrm{meas}),
\end{equation}
up to an unknown overall multiplicative factor. As a result, we can use the same observable $\mathcal{O}_D(z_\mathrm{meas})$ to represent SN data, but now applied to a collection of measurements at redshifts that are much more heavily weighted toward $z_\mathrm{meas} \ll 1$ than were BAO. In this redshift range, the relevant correlations between $w$ and $\Omega_m$ (solid curves in figure \ref{f_toy}) are all predominantly negative, regardless of whether $w$ is shifted at low $z$ or high $z$. This result is again consistent with the empirical constraints we had presented in figure \ref{f_colored}. 

\bibliographystyle{JHEP.bst}
\bibliography{PhantomAll.bib}

\end{document}